\documentclass[preprint,aps]{revtex4}
\usepackage{graphicx}
\usepackage{dcolumn}
\usepackage{bm}
\usepackage{amsmath}
\usepackage{times}

\begin{document}
\preprint{Odd-frequency proximity}
\title{Odd-frequency pairing in normal metal/superconductor junctions}
\author{ Y. Tanaka$^{1,2}$ Y. Tanuma$^{3}$, and A.A. Golubov$^{4}$ }
\affiliation{$^1$Department of Applied Physics, Nagoya University, Nagoya, 464-8603,
Japan \\
$^2$ CREST Japan Science and Technology Cooperation (JST) 464-8603 Japan \\
$^3$ Institute of Physics, Kanagawa University, 3-7-1, Rokkakubashi,
Kanagawa-ku, Yokohama, 221-8686, Japan \\
$^4$ Faculty of Science and Technology, University of Twente, Enshede, The
Netherlands }
\date{\today}

\begin{abstract}
We theoretically study the induced odd-frequency pairing states in ballistic normal metal/superconductor
(N/S) junctions where a superconductor has even-frequency symmetry in the bulk
and a normal metal layer has an arbitrary length.
Using the quasiclassical Green's function formalism, we demonstrate that, quite generally, the pair amplitude
in the junction has an admixture of an odd-frequency component due to the breakdown of translational
invariance near the N/S interface where the pair potential acquires spatial dependence.
%
If a superconductor has even-parity pair potential (spin-singlet $s$-wave state),
the odd-frequency pairing component with odd-parity is induced near the N/S interface, while
in the case of odd-parity pair potential (spin-triplet $p_{x}$-wave or spin-singlet $d_{xy}$-wave)
the odd-frequency component with even-parity is generated. We show that in conventional $s$-wave junctions,
the amplitude of the odd-frequency pairing state is strongest in the case of fully transparent N/S interface
and is enhanced at energies corresponding to the peaks in the local density of states (LDOS).
In $p_x$- and $d_{xy}$-wave junctions, the amplitude of the odd-frequency component on the
S side of the N/S interface is enhanced at zero energy where the midgap Andreev resonant state (MARS)
appears due to the sign change of the pair potential. The odd-frequency component
extends into the N region and exceeds the even-frequency component at energies corresponding to the
LDOS peak positions, including the MARS. At the edge of the N region the odd-frequency component is non-zero
while the even-frequency one vanishes. We show that the concept of odd-frequency pairing is a useful tool
to interpret a number of phenomena in non-uniform superconducting systems, like McMillan-Rowell
and midgap Andreev resonance states.
\end{abstract}

\maketitle




%
%

%




\section{Introduction}

Odd-frequency superconducting pairing state, characterized by pair amplitude
which is an odd function of energy or Matsubara frequency, was first predicted by Berezinskii
\cite{Berezinskii} and has been attracted a lot of interest recently.
Although the existence of odd-frequency pairing in bulk uniform systems is not fully established yet
\cite{Odd1,Odd2,Odd3,Odd4,Odd5}, there is a number of proposals to realize it in
superconducting junctions.
The realization of the odd-frequency pairing state without
finite pair potential was proposed by Bergeret,
Volkov and Efetov in Ref.~\onlinecite{Efetov1}
in ferromagnet/superconductor heterostructures with inhomogeneous magnetization
and several related works have been presented up to now
\cite{Efetov2,Kadigrob,expt}.
In particular, it was predicted that the local density of states
(LDOS) in the ferromagnet is enhanced in the presence of the odd-frequency
pairing \cite{Yokoyama}.

Recently, it was shown that the odd-frequency pairing state is possible
even without magnetic ordering. Two of the present authors predicted that the
odd-frequency pair amplitude can be induced in a diffusive normal metal
attached to a spin-triplet superconductor \cite{Golubov2007}.
According to this study, the origin of the anomalous proximity effect specific to
spin-triplet $p$-wave superconductor junctions \cite{proximityp} is the
realization of the odd-frequency pairing state in the diffusive normal
metal. It is also clarified that the penetration of the midgap Andreev
resonant state (MARS) \cite{ZES,TK95} into the diffusive normal metal
is the manifestation of the existence of the odd-frequency spin-triplet $s$%
-wave superconducting state. The MARS is the well-known resonant state
specific to unconventional superconductors with sign change of the pair
potential on the Fermi surface and was observed experimentally in various
materials \cite{ExpU}.

Furthermore, it was predicted very recently \cite{Ueda,Eschrig2007}
that due to spatial variation
of the pair potential near a N/S interface \cite{Spatial},
the odd-frequency pairing state
can be induced  even in a conventional ballistic N/S system without spin-triplet ordering.
By studying infinite normal metal/infinite superconductor
(N/S) junctions, it was shown that, quite generally, the
spin-singlet even-parity (spin-triplet odd-parity) pair potential in the
superconductor induces the odd-frequency pairing component with spin-singlet
odd-parity (spin-triplet even-parity) near the N/S interface \cite{Ueda}.
The magnitude of the induced odd-frequency component is enhanced in the presence of the MARS
due to the sign change of the anisotropic pair potential at the interface.
In Ref.~\onlinecite{Ueda},
only the generation of the odd-frequency component at
the S side of the N/S interface was studied
by two of the authors.
Therefore the question remains how this component extends into the N region and
how it manifests itself in the properties of the normal metal.
In a semi-infinite ballistic normal metal  attached to a superconductor, the LDOS normalized
by its value in the normal state is always unity.
This well-known property of LDOS is due to the absence of interference between
electrons and Andreev reflected holes in a semi-infinite N metal. In this case, the LDOS
cannot be used to characterize the superconducting correlations in a normal metal.
Thus, in order to understand manifestations of the induced odd-frequency pairing state
in the N metal in a much more clear way, it is necessary to study junctions with finite length of the N region.

In the present paper, using the quasiclassical Green's function theory, we
study the pair amplitude and the LDOS at the N/S interface when the N region
has finite thickness $L$. The spatial dependence of the pair potential is
determined self-consistently. For the convenience of the actual numerical
calculation, we have used the boundary condition in the Ricatti
parametrization \cite{Ricatti}. The superconductor is assumed to have the
conventional even-frequency pairing state in the bulk, being in the
spin-singlet even-parity state ($s$-wave or $d_{xy}$-wave symmetry) or in
the spin-triplet odd-parity state ($p_{x}$-wave symmetry). We show that,
quite generally, the spatial variation of the pair potential and the
proximity effect lead to the generation of the odd-frequency component near
the N/S interface and on the N side. Moreover, when the superconductor is in
the even-parity (odd-parity) state, the resulting odd-frequency component is
odd-parity (even-parity) in order to conserve the spin component. In the
absence of the MARS, like in the case of spin-singlet $s$-wave junctions,
the magnitude of the odd-frequency component of the pair amplitude is
suppressed when the transmission coefficient through the interface decreases.
The resulting odd-frequency pair amplitude has its maximum
value at the interface. At the edge of the N region, the odd-frequency
component is always absent as well as in the S region far away from the
interface. The LDOS is suppressed around $\varepsilon=0$, where $\varepsilon$
is the quasiparticle energy measured from the Fermi level. For large
magnitude of $L$, the resulting LDOS has an oscillatory $\varepsilon$
dependence. The amplitude of the odd-frequency pair amplitude can exceed
that of the even-frequency one at some $\varepsilon$ values. For the case of
spin-triplet $p_{x}$-wave and spin-singlet $d_{xy}$-wave junctions, the
amplitude of the odd-frequency component at the S side of the N/S interface
is much larger than that of the even-frequency pair amplitude. This is
due to the fact that the presence of the MARS at the interface \cite{TK95}
enhances the amplitude of the odd-frequency paring state as shown in Ref.~
\onlinecite{Ueda}. At the edge of the N region, the even-frequency component is
always absent and only the odd-frequency component is nonzero.
At $\varepsilon =0$ the resulting odd-frequency component always exceeds the
even-frequency one.

The organization of the present paper is as follows. In section 2, we
introduce the model and the quasiclassical Green's function formalism. In
section 3, the results of the numerical calculations are discussed for the case
of spin-singlet $s$-wave, spin-triplet $p_{x}$-wave and spin-singlet $d_{xy}$-wave junctions.
In section 4, the conclusions and outlook are presented.

\section{Model and Formulation}
In the following, we consider a N/S junction as the simplest example of
non-uniform superconducting system without impurity scattering. Both cases
of spin-triplet odd-parity and spin-singlet even-parity symmetries are
considered in the superconductor. In the spin-triplet superconductor we
choose $S_{z}=0$ for simplicity. We assume a thin insulating barrier
located at the N/S interface ($x=0$) with N ($-L<x<0)$ and S ($x>0$) modeled
by a delta function $H\delta (x)$, where $H $ is the
strength of the delta function potential. The length of the normal region is
$L$. The reflection coefficient of the junction for the quasiparticle for
the injection angle $\theta $ is given by $R=Z^{2}/(Z^{2}+4\cos ^{2}\theta )$
with $Z=2H/v_{F}$, where $\theta $ $(-\pi /2<\theta <\pi /2)$ is measured
from the normal to the interface and $v_{F}$ is the Fermi velocity.

The quasiclassical Green's functions \cite{Quasi} in a normal metal (N) and
a superconductor (S) in the Matsubara frequency representation are
parameterized as
\begin{equation}
\hat{g}_{\pm }^{(i)}=f_{1\pm }^{(i)} \hat{\tau}_{1}+f_{2\pm }^{(i)}\hat{\tau}%
_{2} +g_{\pm }^{(i)}\hat{\tau}_{3},\ \ (\hat{g}_{\pm }^{(i)})^{2}=\hat{1}
\end{equation}%
where the subscript $i(=N,S)$ refer to N and S, respectively. Here, $\hat{%
\tau}_{j}$ ($j=1,2,3$) are Pauli matrices and $\hat{1}$ is a unit matrix. The
subscript $+(-)$ denotes the left (right) going quasiparticles \cite{Serene}%
. Functions $\hat{g}_{\pm }^{(i)}$ satisfy the Eilenberger equation~\cite%
{Eilen}
\begin{equation}
iv_{Fx}\hat{g}_{\pm}^{(i)} = \mp[\hat{H}_{\pm},\hat{g}_{\pm}^{(i)}]
\end{equation}
with
\begin{equation*}
\hat{H}_{\pm}=i\omega_{n}\tau_{3} + i\bar{\Delta}_{\pm}(x)\tau_{2}.
\end{equation*}
Here $v_{Fx}$ is the $x$ component of the Fermi velocity, $\omega
_{n}=2\pi T(n+1/2)$ is the Matsubara frequency, $n$ is an integer number and $T$
is temperature. $\bar{\Delta}_{+}(x)$ ($\bar{\Delta}_{-}(x)$) is the
effective pair potential for left (right) going quasiparticles. In the N
region, $\bar{\Delta}_{\pm}(x)$ is set to zero due to the absence of a
pairing interaction in the N metal. The above Green's
functions can be expressed as
\begin{align}
f_{1\pm }^{(i)} &=\pm i(F_{\pm }^{(i)}+D_{\pm }^{(i)})
/(1-D_{\pm }^{(i)}F_{\pm }^{(i)}), \\
f_{2\pm}^{(i)}&=-(F_{\pm }^{(i)}-D_{\pm }^{(i)})
/(1-D_{\pm }^{(i)}F_{\pm }^{(i)}),  \notag \\
g_{\pm}^{(i)}&=(1+D_{\pm }^{(i)}F_{\pm }^{(i)})
/(1-D_{\pm }^{(i)}F_{\pm }^{(i)}),  \notag
\end{align}
where $D_{\pm }^{(i)}(x)$ and $F_{\pm }^{(i)}(x)$
satisfy the Ricatti equations \cite%
{Ricatti} in the N region
\begin{align}
v_{Fx}\partial _{x}D_{\pm }^{(N)}(x) &=-2\omega_{n}D_{\pm }^{(N)}(x)  \label{eq.1a} \\
v_{Fx}\partial _{x}F_{\pm }^{(N)}(x) &=2\omega_{n}F_{\pm }^{(N)}(x),  \label{eq.1b}
\end{align}
and in the S region,
\begin{align}
v_{Fx}\partial _{x}D_{\pm }^{(S)}(x)
&=-\bar{\Delta}_{\pm }(x)[1-(D_{\pm
}^{(S)}(x))^{2}] +2\omega_{n}D_{\pm }^{(S)}(x)  \label{eq.1c} \\
v_{Fx}\partial _{x}F_{\pm }^{(S)}(x) &=-\bar{\Delta}_{\pm }(x)
[1-(F_{\pm}^{(S)}(x))^{2}] -2\omega_{n}F_{\pm }^{(S)}(x),  \label{eq.1d}
\end{align}
respectively.

The boundary conditions at the edge of N region, $x=-L$, have the form
\begin{equation}
F_{+}^{(N)}(-L)=-D_{-}^{(N)}(-L),
\ \ F_{-}^{(N)}(-L)=-D_{+}^{(N)}(-L)
\end{equation}%
The boundary conditions at the N/S interface, $x=0$, are
\begin{equation}
F_{\pm }^{(S)}(0)
=-\frac{(1-R)D_{\pm }^{(N)}(0)+[R+D_{+}^{(N)}(0)D_{-}^{(N)}(0)]
D_{\mp}^{(S)}(0)}{%
[1+RD_{+}^{(N)}(0)D_{-}^{(N)}(0)]
+(1-R)D_{\mp}^{(N)}(0)D_{\mp}^{(S)}(0)}
\end{equation}%
and
\begin{equation}
F_{\pm }^{(N)}(0)
=-\frac{(1-R)D_{\pm }^{(S)}(0)+
[R+D_{+}^{(S)}(0)D_{-}^{(S)}(0)]D_{\mp}^{(N)}(0)}
{[1+RD_{+}^{(S)}(0)D_{-}^{(S)}(0)]
+(1-R)D_{\mp}^{(N)}(0)D_{\mp }^{(S)}(0)},
\end{equation}%
where $R$ is the reflection coefficient at the interface. Since there is no
pair potential in the N region, the solutions for the spatial dependence of
above functions can be easily found
\begin{equation*}
\begin{split}
D_{-}^{(N)}(x)=-At^{-1},& \quad D_{+}^{(N)}(x)=-Bt^{-1} \\
F_{+}^{(N)}(x)=At,& \quad F_{-}^{(N)}(x)=Bt,
\end{split}%
\end{equation*}%
%
%
with $t=\exp [(x+L)/\xi ]$ and $\xi =\hbar v_{Fx}/2|\omega _{n}|$. The
constants $A$ and $B$ are given by
\begin{equation}
A=\frac{-2(1-R)D_{+}^{(S)}(0)t_{0}}{\Lambda +\sqrt{\Lambda
^{2}+4(1-R)^{2}t_{0}^{2}D_{+}^{(S)}(0)D_{-}^{(S)}(0)}}
\end{equation}%
\begin{equation}
B=\frac{-2(1-R)D_{-}^{(S)}(0)t_{0}}
{\Lambda +\sqrt{\Lambda^{2}
+4(1-R)^{2}t_{0}^{2}D_{+}^{(S)}(0)D_{-}^{(S)}(0)}}
\end{equation}%
with
\begin{equation*}
\Lambda =t_{0}^{2}(1+RD_{+}^{(S)}(0)D_{-}^{(S)}(0))
-(R+D_{+}^{(S)}(0)D_{-}^{(S)}(0))
\end{equation*}%
with $t_{0}=\exp (L/\xi )$.
After simple manipulation, we obtain$f_{1\pm}^{(N)}$,
$f_{2\pm}^{(N)}$ and $g_{\pm}^{(N)}$
\begin{gather}
\begin{split}
\label{f12}
f_{1+}^{(N)}=-i(At-B/t)/(1+AB), \quad
f_{1-}^{(N)}=i(Bt-A/t)/(1+AB),
\\
f_{2+}^{(N)}=-(At +B/t)/(1+AB), \quad
f_{2-}^{(N)}=-(Bt + A/t)/(1+AB),
\end{split}
\\
\begin{split}
\label{LDOS}
g_{+}^{(N)}=g_{-}^{(N)}=(1-AB)/(1+AB).
\end{split}
\end{gather}
Note that as follows from Eq. (\ref{LDOS}), functions $g_{+}^{(N)}$ and
$g_{-}^{(N)}$
are spatially-independent.

Here, we consider the situation without mixing of different symmetry
channels for the pair potential. Then the pair potential $\bar{\Delta}_{\pm
}(x)$ is expressed by
\begin{equation}
\bar{\Delta}_{\pm}(x)=\Delta (x)\Phi _{\pm }(\theta )\Theta(x)
\end{equation}
with the form factor $\Phi_{\pm }(\theta )$ given by $\Phi_{\pm }(\theta )=1$%
, $\pm \sin 2\theta $, and $\pm \cos \theta $ for $s$-wave, $d_{xy}$-wave,
and $p_{x}$-wave superconductors, respectively. The pair potential $%
\Delta(x) $ is determined by the self-consistent equation 
\begin{equation}
\Delta (x)=\frac{2 T}{\mathrm{log}\frac{T}{T_{C}}+\displaystyle\sum_{n\geq 1}%
\frac{1}{n-\frac{1}{2}}}\displaystyle\sum_{n\geq 0}\int_{-\pi /2}^{\pi
/2}d\theta G(\theta )f_{2+}
\end{equation}%
with $G(\theta )=1$ for $s$-wave case and $G(\theta )=2\Phi (\theta )$ for
other cases, respectively \cite{Matsumoto}. $T_{C}$ is the transition
temperature of the superconductor. %
The condition in the bulk is $\Delta (\infty )=\Delta_{0} $. Since the pair
potential $\bar{\Delta}(x)$ is a real quantity, the resulting $f_{1\pm }$ is
an imaginary quantity and $f_{2\pm }$ is a real one.

Before performing actual numerical calculations, we now discuss general
properties of the pair amplitude. In the following, we explicitly write $%
f_{1\pm }^{(i)}=f_{1\pm }^{(i)}(\omega _{n},\theta )$, $f_{2\pm
}^{(i)}=f_{2\pm }^{(i)}(\omega _{n},\theta )$, $F_{\pm }^{(i)}=F_{\pm
}^{(i)}(\omega _{n},\theta )$ and $D_{\pm }^{(i)}=D_{\pm }^{(i)}(\omega
_{n},\theta )$. For the limit $x=\infty $, we obtain
\begin{equation}
f_{1\pm }^{(S)}(\omega _{n},\theta )=0,\quad f_{2\pm }^{(S)}(\omega
_{n},\theta )=\frac{\Delta_{0} \Phi _{\pm }(\theta )}{\sqrt{\omega
_{n}^{2}+\Delta _{0}^{2}\Phi _{\pm }^{2}(\theta _{\pm })}}.
\end{equation}%
Note that $f_{1\pm }^{(i)}(\omega _{n},\theta )$ becomes finite due to the
spatial variation of the pair potential and it does not exist in the bulk.
From Eqs. (\ref{eq.1a}) and (\ref{eq.1b}), we can show that $D_{\pm
}^{(i)}(-\omega _{n},\theta )=1/D_{\pm }^{(i)}(\omega _{n},\theta )$ and
$F_{\pm }^{(i)}(-\omega _{n},\theta )
=1/F_{\pm }^{(i)}(\omega _{n},\theta )$.
After simple manipulation, we obtain
\begin{equation}
f_{1\pm }^{(i)}(\omega _{n},\theta )=-f_{1\pm }^{(i)}(-\omega _{n},\theta
),\quad f_{2\pm }^{(i)}(\omega _{n},\theta )=f_{2\pm }^{(i)}(-\omega
_{n},\theta ),
\end{equation}%
for any $x$. It is remarkable that functions $f_{1\pm }^{(i)}(\omega
_{n},\theta )$ and $f_{2\pm }^{(i)}(\omega _{n},\theta )$ correspond to
odd-frequency and even-frequency components of the pair amplitude,
respectively \cite{Efetov2,Ueda}. Function $f_{1\pm }^{(1)}(\omega
_{n},\theta )$ describes the odd-frequency component of the pair amplitude
penetrating from the superconductor.

Next, we discuss the parity of these pair amplitudes. The even-parity
(odd-parity) pair amplitude should satisfy the following relation $f_{j\pm
}^{(i)}(\omega_{n},\theta ) =f_{j\mp }^{(i)}(\omega _{n},-\theta )$ [$%
f_{j\pm }^{(i)}(\omega_{n},\theta ) =-f_{j\mp }^{(i)}(\omega _{n},-\theta )$%
], with $j=1,2$. For an even-parity (odd-parity) superconductor, $\Phi _{\pm
}(-\theta )=\Phi _{\mp }(\theta)$ [$\Phi _{\pm }(-\theta )=-\Phi _{\mp
}(\theta )$]. Then, we can show that for the even-parity case
\begin{equation}
D_{\pm }^{(i)}(-\theta )=D_{\mp }^{(i)}(\theta ), \quad F_{\pm
}^{(i)}(-\theta )=F_{\mp}^{(i)}(\theta )
\end{equation}
and for the odd-parity case
\begin{equation*}
D_{\pm }^{(i)}(-\theta )=-D_{\mp }^{(i)}(\theta ), \quad F_{\pm
}^{(i)}(-\theta )=-F_{\mp }^{(i)}(\theta )
\end{equation*}
respectively.

The resulting $f_{1\pm }^{(i)}(\omega _{n},\theta )$ and $%
f_{2\pm}^{(i)}(\omega _{n},\theta )$ satisfy
\begin{gather}
\begin{split}
f_{1\pm }^{(i)}(\omega _{n},\theta ) =-f_{1\mp}^{(i)}(\omega _{n},-\theta ),
\\
f_{2\pm }^{(i)}(\omega _{n},\theta ) =f_{2\mp}^{(i)}(\omega _{n},-\theta ),
\end{split}%
\end{gather}
for an even-parity superconductor and
\begin{gather}
\begin{split}
f_{1\pm}^{(i)}(\omega _{n},\theta ) =f_{1\mp}^{(i)}(\omega _{n},-\theta ), \\
f_{2\pm}^{(i)}(\omega _{n},\theta ) =-f_{2\mp }^{(i)}(\omega _{n},-\theta ),
\end{split}%
\end{gather}
for an odd-parity superconductor, respectively \cite{Ueda}. Note that the
parity of the odd-frequency component $f_{1\pm}^{(i)}(\omega _{n},\theta )$
is always different from that in the bulk superconductor.

As shown above, the odd-frequency component $f_{1\pm}^{(i)}(\omega
_{n},\theta )$ is purely an imaginary quantity. The underlying physics behind
this formal property is the follows. Due to the breakdown of translational
invariance near the N/S interface, the pair potential $\bar{\Delta}(x)$
acquires a spatial dependence which leads to the coupling between even-parity
and odd-parity states. Since the bulk pair potential has an
even-frequency symmetry, the Fermi-Dirac statistics requires that the order
parameter component induced near the interface should be odd in frequency.
The phase of the induced pair amplitude undergoes a $\pi/2$ shift from that in the bulk S
thus removing internal phase shift between the even- and odd-frequency components
and making the interface-induced state compatible with the time reversal invariance.
As a result, function $f_{1\pm}^{(i)}(\omega _{n},\theta )$ becomes a
purely imaginary quantity \cite{Ueda}.

Let us now focus on the values of the pair amplitudes at the edge of N
region (at $x=-L$). We concentrate on two extreme cases with (I) $\Phi
_{+}(\theta)=\Phi _{-}(\theta )$ and (II) $\Phi _{+}(\theta )=-\Phi
_{-}(\theta)$. In the case (I), the MARS is absent since there is no
sign change of the pair potential felt by the quasiparticle at the
interface. Then the relation $D_{+}^{(N)}=D_{-}^{(N)}$ holds. On the other hand,
in the case (II), the MARS is generated near the interface due to the
sign change of the pair potential and the relation $D_{+}^{(N)}=-D_{-}^{(N)}$ is
satisfied \cite{TK95}. At the edge $x=-L$, it is easy to show that $%
F_{\pm}^{(N)}=-D_{\pm}^{(N)}$ for the former case and $F_{\pm}^{(N)}=D_{%
\pm}^{(N)}$ for the latter one. As a result, $f_{1\pm }^{(N)}=0$ for the
case (I) and $f_{2\pm }^{(N)}=0$ for the case (II), respectively. Thus we
can conclude that in the absence of the MARS only the even-frequency pairing
component exists at $x=-L$, while in the presence of the MARS only the odd-frequency one.

In order to understand the angular dependence of the pair amplitude in a
more detail, we define $\hat{f}_{1}^{(i)}$ and $\hat{f}_{2}^{(i)}$ for $-\pi
/2<\theta<3\pi /2$ with
$\hat{f}_{1(2)}^{(i)}=f_{1(2)+}^{(i)}(\theta )$ for $-\pi
/2<\theta <\pi/2$ and
$\hat{f}_{1(2)}^{(i)}=f_{1(2)-}^{(i)}(\pi -\theta )$ for
$\pi/2<\theta <3\pi /2$. We decompose $\hat{f}_{1(2)}^{(i)}$ into various
angular momentum component as follows,
\begin{equation}
\displaystyle\hat{f}_{1(2)}^{(i)} =
\sum_{m}S_{m}^{(1(2))} \sin (m\theta )+
\sum_{m}C_{m}^{(1(2))}\cos (m\theta )
\end{equation}
with $m=2l+1$ for odd-parity case and $m=2l$ for even-parity case with
integer $l\geq 0$, where $l$ is the quantum number of the angular momentum.
Here, $C_{m}^{(1(2))}$ and $S_{m}^{(1(2))}$ 
are defined for all $x$. It is
straightforward to show that the only nonzero components are (1) $%
C_{2l}^{(2)}$ and $C_{2l+1}^{(1)}$ for even-parity superconductor without
sign change at the interface ($i.e.$, $s$-wave or $d_{x^{2}-y^{2}}$-wave),
(2) $S_{2l+2}^{(2)}$ and $S_{2l+1}^{(1)}$ for $d_{xy}$-wave, (3) $%
C_{2l+1}^{(2)}$ and $C_{2l}^{(1)}$ for $p_{x}$-wave, and (4) $S_{2l+1}^{(2)}$
and $S_{2l}^{(1)}$ for $p_{y}$-wave junctions, respectively. The allowed
angular momenta for odd-frequency components are $2l+1$, $2l+1$, $2l$, and $%
2l+2$ corresponding to each of the above four cases. 

In order to get better insight into the spectral property of the
odd-frequency pair amplitude, we perform an analytical continuation from the
Matsubara frequency $\omega _{n}$ to the quasiparticle energy $\varepsilon $
measured from the chemical potential.
The retarded Green's function corresponding to Eq. (1) is defined as
$\hat{g}_{\pm }^{(i)R}=f_{1\pm}^{(i)R}\hat{\tau} _{1} +f_{2\pm }^{(i)R}
\hat{\tau}_{2} +g_{\pm }^{(i)R}\hat{\tau}_{3}$.
One can show that
$f_{1\pm}^{(i)R}(-\varepsilon )=-[f_{1\pm}^{(i)R}(\varepsilon )]^{\ast }$,
$f_{2\pm }^{(i)R}(-\varepsilon )=[f_{2\pm}^{(i)R}(\varepsilon )]^{\ast }$,
and
$g_{\pm}^{(i)R}(-\varepsilon)=[g_{\pm}^{(i)R}(\varepsilon )]^{\ast }$.
The LDOS $\rho (\varepsilon )$ at the N/S interface at $x=0$
normalized to its value in the normal state is given by
\begin{equation}
\rho (\varepsilon )=\int_{-\pi /2}^{\pi /2}d\theta \mathrm{Real} %
\displaystyle( \frac{g_{+}^{(i)R}(\varepsilon) +g_{-}^{(i)R}(\varepsilon)}
{2\pi } \displaystyle)
\end{equation}
In the following, we self-consistently calculate
the spatial dependence of the pair potential and the pair amplitude
in the Matsubara representation.
After that we calculate the spectral properties of pair amplitudes and LDOS.
For actual calculations, we choose spin-singlet $s$-wave,
spin-triplet $p_{x}$-wave and spin-singlet $d_{xy}$-wave state in a
superconductor and fix temperature $T=0.05T_{C}$.
The length of the normal region $L$
is measured in units of $L_{0}=v_{F}/2\pi T_{C}$.

\section{Results}

\subsection{$s$-wave pair potential}

\begin{figure}[tb]
\begin{center}
\scalebox{0.8}{
\includegraphics[width=20cm,clip]{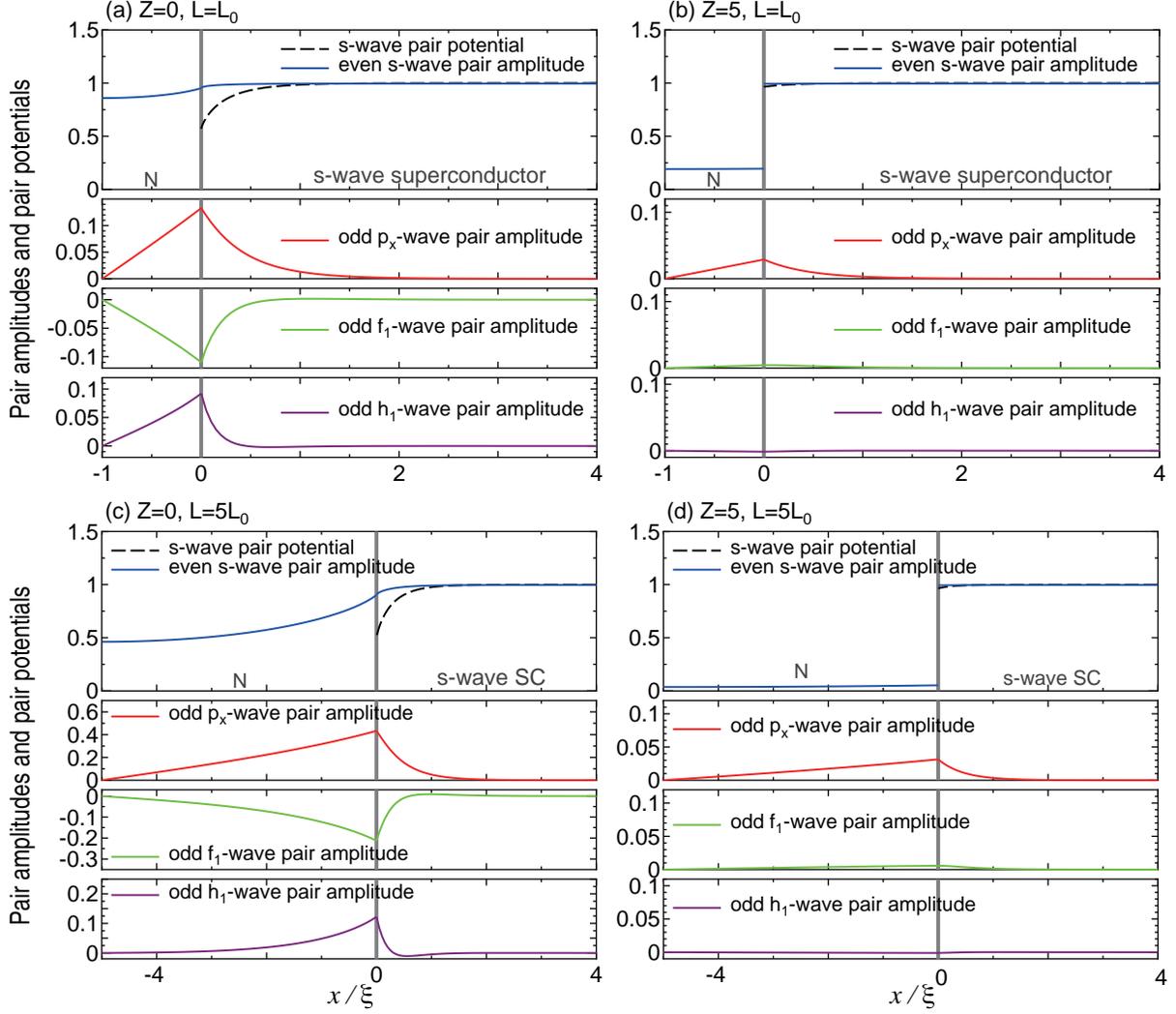}
}
\end{center}
\caption{(Color online) Spatial dependence of the normalized pair potential,
even-frequency pair amplitude and odd-frequency components of the pair
amplitude for $s$-wave superconductor junctions. Here, we choose $\protect\xi%
=v_{F}/\Delta_{0}$ in the S region $(x>0)$ and $\protect\xi=L_{0}=v_{F}/2%
\protect\pi T_{C}$ in the N region. The pair amplitudes $C_{0}^{(2)}$,
$C_{1}^{(1)}$, $C_{3}^{(1)}$, and $C_{5}^{(1)}$ are denoted as even $s$-wave,
odd $p_{x}$-wave, odd $f_{1}$-wave, and odd $h_{1}$-wave pair
amplitudes. (a) $Z=0$, $L=L_{0}$, (b) $Z=5$, $L=L_{0}$, (c) $Z=0$, $L=5L_{0}$%
, and (d) $Z=5$, $L=5L_{0}$, respectively. }
\label{fig:1}
\end{figure}

First we focus on the $s$-wave superconductor junctions as shown in Fig.~\ref%
{fig:1}. By changing the length $L$ of the N region and the transparency at
the interface, we calculate the spatial dependence of the pair potential and
the pair amplitudes in the Matsubara frequency representation. We only
concentrate on the lowest angular momentum of the even-frequency pair
amplitude $C_{0}^{(2)}$. As regards the odd-frequency pair amplitudes, we
focus on the $C_{1}^{(1)}$, $C_{3}^{(1)}$ and $C_{5}^{(1)}$ components which
all have odd-parity and depend on $\theta $ as $\cos \theta $, $\cos 3\theta $
and $\cos 5\theta $, respectively, and correspond to $p_{x} $-wave,
$f_{1}$-wave and $h_{1}$-wave components shown in Fig. 1.
In all cases, even-frequency component
is constant in the S region far away from the interface and the
corresponding odd-frequency components are absent. The $s$-wave pair
potential is suppressed for the fully transparent case ($Z=0$), while it is
almost constant for low transparent case ($Z=5$). It does not penetrate into
the N region due to the absence of the attractive interaction in the N metal.
On the other hand, in all considered cases the spatial variation of the
even-frequency $s$-wave pair amplitude is rather weak in the S region, while
in the N region it is strong for $Z=0$ and is reduced for $Z=5$
since the proximity effect is weaker in the latter case. The odd-frequency
component always vanishes at $x=-L$ and does not have a jump at the N/S
interface even for nonzero $Z$. Its amplitude is strongly enhanced near the
N/S interface especially for fully transparent junctions. Note that
not only the $p_{x}$-wave but also $f_{1}$-wave and $h_{1}$%
-wave have sufficiently large magnitudes as shown in Figs.~\ref{fig:1}(a)
and \ref{fig:1}(c). With the decrease of the transparency of the N/S
interface, the odd-frequency components are
suppressed as shown in Figs.~\ref{fig:1}(b) and \ref{fig:1}(d).

\begin{figure}[tb]
\begin{center}
\scalebox{0.8}{
\includegraphics[width=9cm,clip]{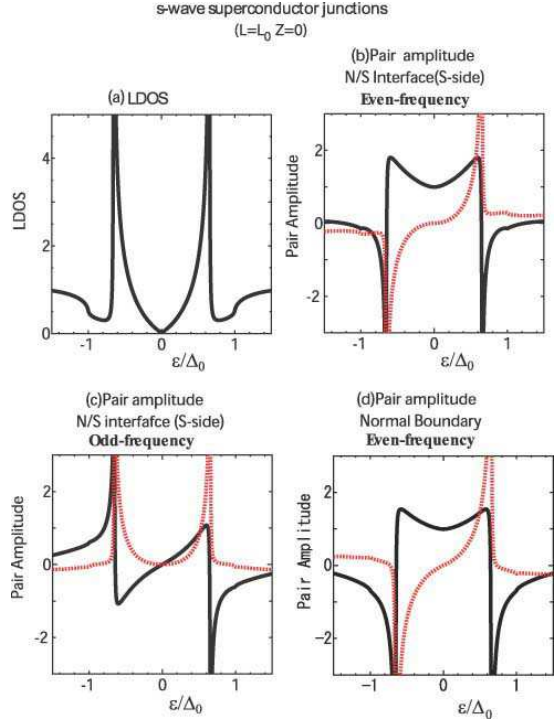}
}
\end{center}
\caption{(Color online) Energy dependence of the LDOS and the pair amplitudes
in $s$-wave junctions with $L=L_{0}$ and $Z=0$. (a) The LDOS normalized by its value in the normal
state. The solid line: LDOS on the S side of the N/S interface, the dotted
line: LDOS in the N region. Energy dependence of the real (solid line) and
imaginary (dotted line) part of (b) even-frequency $s$-wave pair amplitude on the S
side of the N/S interface, (c) odd-frequency $p_{x}$-wave pair amplitude on
the S side of the N/S interface and (d) even-frequency $s$-wave pair amplitude
at the edge of the N region. }
\label{fig:2}
\end{figure}

In order to understand the proximity effect in more detail, we look at
the resulting LDOS and the spectral properties of pair amplitudes in real
energy $\varepsilon $. We focus on the even-frequency $s$-wave component of
the pair amplitude $C_{0}^{(1)}$ and on the odd-frequency $p_{x}$-wave pair
amplitude $C_{1}^{(1)}$ on the S side of the N/S interface and on the edge of the N region.
As follows from Eq. (\ref{LDOS}), the LDOS is independent of the coordinate $x$ in the N.
For $Z=0$ and $L=L_{0}$, the LDOS has a V-shaped structure. There is no
jump of the value of the LDOS at the N/S interface. The even-frequency pair
amplitude at the N/S boundary on the S side is shown in Fig.~\ref{fig:2}(b).
Its real part is an even function of $\varepsilon $ while its imaginary part
is an odd function of $\varepsilon $. The corresponding odd-frequency one is
plotted in Fig. \ref{fig:2}(c). In contrast to the even-frequency component [Fig.~\ref{fig:2}(b)],
the real (imaginary) part of the pair amplitude is an odd (even) function of
$\varepsilon $.
The pair amplitude is enhanced around $\varepsilon \sim \pm
0.6\Delta _{0}$ where the LDOS have a peak. At the N/S boundary ($x=-L$),
only the even-frequency component exists. The line shape of the pair
amplitude shown in Fig.~\ref{fig:2}(d) is similar to that
in Fig.~\ref{fig:2}(b).

\begin{figure}[tb]
\begin{center}
\scalebox{0.8}{
\includegraphics[width=9cm,clip]{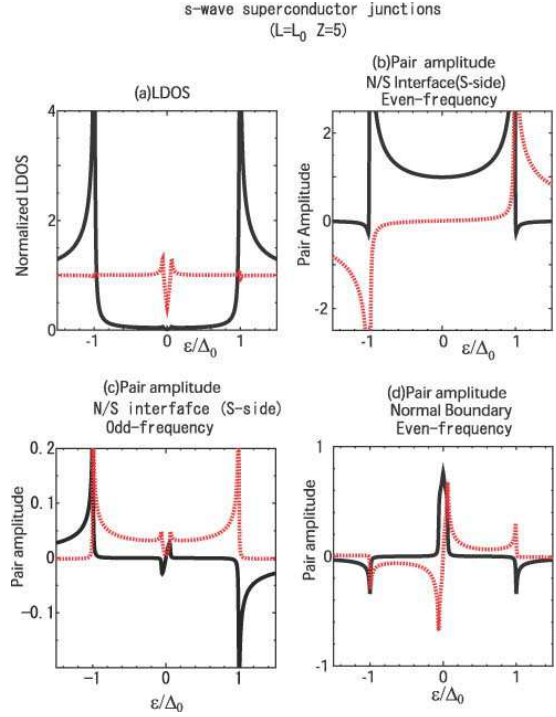}
}
\end{center}
\caption{(Color online) Same as Fig.~\protect\ref{fig:2}, but with $L=L_{0}$ and $Z=5$. }
\label{fig:3}
\end{figure}

In Fig.~\ref{fig:3}, the corresponding plots for $L=L_{0}$ and $Z=5$ are
shown. The LDOS on the S side of the N/S interface has a U-shaped DOS
similar to bulk DOS. On the other hand, in the N region, the LDOS has a
different value due to the discontinuity at the N/S interface. The LDOS
in N also exhibits the minigap structure which scales with the interface
transparency, in accordance with the well-known McMillan model of proximity
effect in conventional superconducting N/S junctions~\cite{McMillan}.
The magnitude of the real part of the even-frequency component on the S
side of the N/S interface exceeds the magnitude of the imaginary
part as seen from
Fig.~\ref{fig:3}(b) for $|\varepsilon |<\Delta _{0}$. The magnitude of the
odd-frequency part is small as compared to that of the even-frequency one.
As seen from Fig.~\ref{fig:3}(c), the real part of the odd-frequency
component has a minigap structure and the imaginary part has a dip and peak
structure in contrast to the case of the even-frequency one [see
Fig.~\ref{fig:3}(b)]. At the N/S boundary ($x=-L$), only the even-frequency
component exists. The real part of the even-frequency component at $x=-L$
has a peak around $\varepsilon =0$ [see Fig.~\ref{fig:3}(d)]. The width of
this peak is of the same order as that of the dip of LDOS. As compared to
the corresponding case of $Z=0$, the proximity effect in the N region is
only essential at low energy $\varepsilon $.

\begin{figure}[tb]
\begin{center}
\scalebox{0.8}{
\includegraphics[width=9cm,clip]{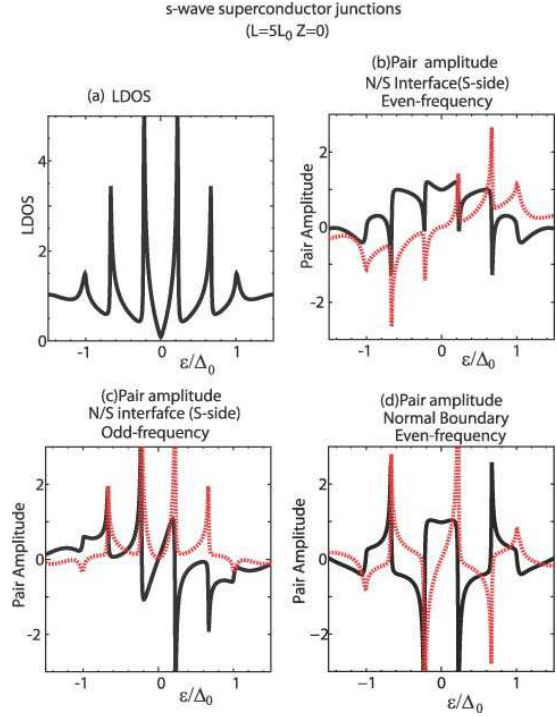}
}
\end{center}
\caption{(Color online) Same as Fig.~\protect\ref{fig:2}, but with $L=5L_{0}$ and $Z=0$. }
\label{fig:4}
\end{figure}

It is also interesting to consider the case of large width of the N region.
Here, we concentrate on the situation when the N/S/ interface is fully
transparent ($Z=0$) and $L=5L_{0}$. In this case the LDOS in the N region and
at the N/S interface coincide with each other as seen from Fig.~\ref{fig:4}.
The LDOS has multiple peaks due to the existence of the multi-sub
gap structures due to electron-hole interference effects in the N region
\cite{Rowell}.
The amplitudes of the corresponding even-frequency and
odd-frequency components are enhanced at energies $\varepsilon $
corresponding to the LDOS peak positions, while the ratio of this components
depends on energy and location in the N region.
To clarify this point much more clearly, we concentrate on the ratio of the
odd- and even-frequency components in the N region.
According to Eq. (\ref{f12}), the ratio of the magnitude of the odd-frequency
component $f_{1+}^{(N)}(\varepsilon,\theta)$
to the even-frequency one $f_{2+}^{(N)}(\varepsilon,\theta)$  is
\begin{equation}
\frac{\left\vert f_{1+}^{(N)}(\varepsilon,\theta)\right\vert }{\left\vert
f_{2+}^{(N)}(\varepsilon,\theta)\right\vert }=\frac{\left\vert 1/t-t\right\vert }{\left\vert
1/t+t\right\vert }=\left\vert \tan \left( \frac{2\varepsilon }{v_{Fx}}%
(L+x)\right) \right\vert .  \label{f12_E}
\end{equation}
At the edge of the N region, $x=-L$, the odd-frequency component
vanishes at all energies. On the other hand, very interesting situation
occurs at the N/S interface, $x=0$ as will be shown below.
In Fig. \ref{fig:4-1}, we plot this ratio for $\theta=0$ and $x=0$.
\begin{figure}[tb]
\begin{center}
\scalebox{0.8}{
\includegraphics[width=4.5cm,clip]{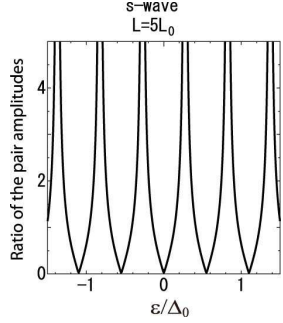}
}
\end{center}
\caption{
Ratio of the pair amplitudes
$f_{1+}^{(N)}(\varepsilon,\theta)/f_{2+}^{(N)}(\varepsilon,\theta)$
on the N-side of the N/S interface in $s$-wave junction as a function of energy $\varepsilon$
for $\theta=0$ and $L=5L_{0}$.}
\label{fig:4-1}
\end{figure}
It is remarkable that at some energies the
amplitude of the odd-frequency pair amplitude exceeds that of the
even-frequency one. \par
Let us clarify the relation between the positions of the bound states and the
above ratio of the odd-to-even pair amplitude.
In the limit $L>>L_{0}$ the bound states are determined by simple relation \cite{Rowell}

\begin{equation}
\varepsilon _{n}=\frac{\pi v_{Fx}}{2L}(n+1/2),\text{ \ \ }n=0,1,2,...
\end{equation}%
Very dramatic situation
occurs at the N/S interface, $x=0$: combining the above two equations, we
obtain that at the LDOS peak positions $\varepsilon =\varepsilon _{n}$
the ratio of the odd-to-even pair amplitude diverges
\begin{equation}
\frac{\left\vert f_{1+}^{(N)}(\varepsilon,\theta)\right\vert }
{\left\vert f_{2+}^{(N)}(\varepsilon,\theta)\right\vert}
=\left\vert \tan \left( \pi /2+\pi n\right) \right\vert =\infty .
\label{f12_sgp}
\end{equation}

That means that at the subgap peak energies the odd-frequency component
dominates over the even-frequency one at the N/S interface.
This is a remarkable property of the
odd-frequency pairing, which makes it relevant to the classical
McMillan-Rowell oscillations in the N/S geometry \cite{Rowell}.
To summarize,
we have shown that the  odd-frequency component is present even in the
standard case of a ballistic N/S system, and it dominates at energies
when the LDOS has subgap peaks.

\subsection{$p_{x}$-wave pair potential}

\begin{figure}[tb]
\begin{center}
\scalebox{0.8}{
\includegraphics[width=20cm,clip]{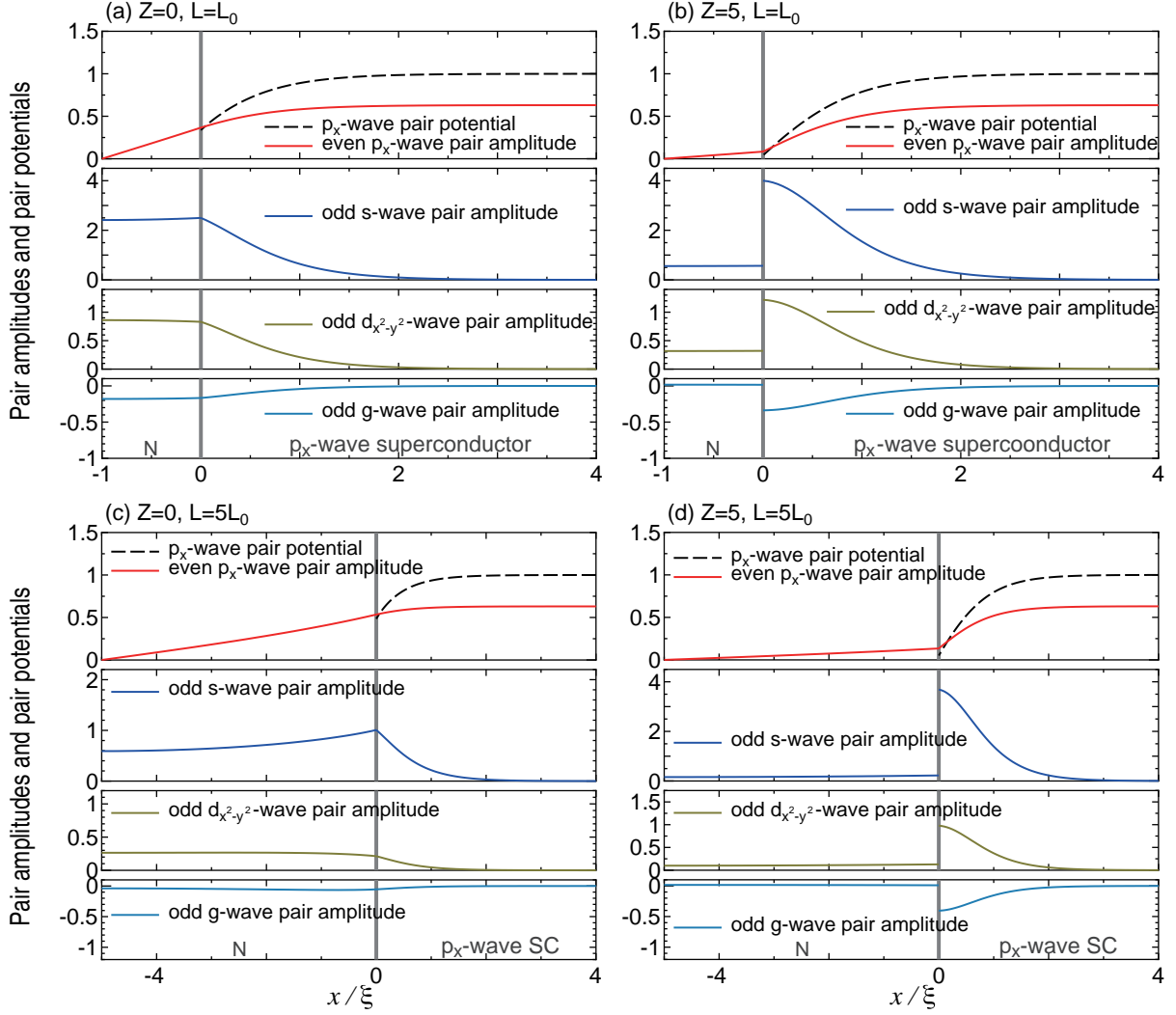}
}
\end{center}
\caption{(Color online) Spatial dependence of the normalized pair potential,
even-frequency and odd-frequency pair amplitudes for $p_{x}$-wave
superconductor
junctions. Here, we choose $\protect\xi=v_{F}/\Delta_{0}$ in the S region $%
(x>0)$ and $\protect\xi=L_{0}=v_{F}/2\protect\pi T_{C}$ in the N region. The
pair amplitudes $C_{1}^{(2)}$, $C_{0}^{(1)}$, $C_{2}^{(1)}$, and $%
C_{4}^{(1)}$ are denoted as even $p_{x}$-wave, odd $s$-wave, 
odd $d_{x^{2}-y^{2}}$-wave, and odd $g$-wave pair amplitudes.
(a) $Z=0$, $L=L_{0}$, (b) $Z=5$, $L=L_{0}$, (c) $Z=0$, $L=5L_{0}$, and (d) $%
Z=5$, $L=5L_{0}$, respectively. }
\label{fig:5}
\end{figure}

Next, we focus on the $p_{x}$-wave superconductor junctions as shown in Fig.~%
\ref{fig:5}. Similar to the case of $s$-wave junctions, by changing the
length of the normal region $L$ and the transparency at the interface, we
calculate the spatial dependence of the pair potential and the pair
amplitudes in the Matsubara frequency representation. We only concentrate on
the lowest angular momentum of the even-frequency pair amplitude $%
C_{1}^{(2)} $. As regards the odd-frequency pair amplitude, we focus on the $%
C_{0}^{(1)}$, $C_{2}^{(1)}$ and $C_{4}^{(1)}$ components where the parity of
the odd-frequency components is even. These functions correspond to $s$-wave,
$d_{x^{2}-y^{2}}$-wave and $g$-wave components in Fig.~\ref{fig:5}, where the
$\theta$ dependencies are given by $1$, $\cos 2\theta$, and $\cos 4\theta$,
respectively. In all cases, even-frequency component is constant in the S
region far away from the interface and the corresponding odd-frequency
components are absent. The $p_{x}$-wave pair potential is reduced at the N/S
interface in all cases. For $Z=5$, the reduction is significant and the
resulting magnitude of the $p_{x}$-wave pair potential is almost zero at the
N/S interface. It does not penetrate into the N region due to the absence of
the attractive interaction in the N metal. The amplitude of the $p_{x}$-wave
even-frequency pair amplitude is reduced towards the N/S interface and
monotonically decreases in the N region. It does not have a jump at the N/S
interface even for nonzero $Z$ and vanishes at the edge of the N region ($%
x=-L $). On the other hand, the odd-frequency component is always nonzero at
$x=-L $ and has a jump at the N/S interface for nonzero $Z$ [see Figs.~\ref%
{fig:5}(b) and \ref{fig:5}(d)]. The amplitude of the odd-frequency component
is strongly enhanced near the S-side of the N/S interface. This enhancement
is much more significant for the low transparent interface with large $Z$
[see Figs.~\ref{fig:5}(b) and \ref{fig:5}(d)]. However, for the presently
chosen Matsubara frequency $\omega_{n}=0.05\pi T_{C}$ it cannot penetrate
into the N region. On the other hand, for $Z=0$, the odd-frequency component
significantly extends into the N region. Note that not only the $s$-wave but
also $d_{x^{2}-y^{2}}$-wave and $g$-wave components have sufficiently large
magnitudes as shown in Figs.~\ref{fig:5}(a) and \ref{fig:5}(c). These pair
amplitudes are almost constant in the N region.

\begin{figure}[tb]
\begin{center}
\scalebox{0.8}{
\includegraphics[width=9cm,clip]{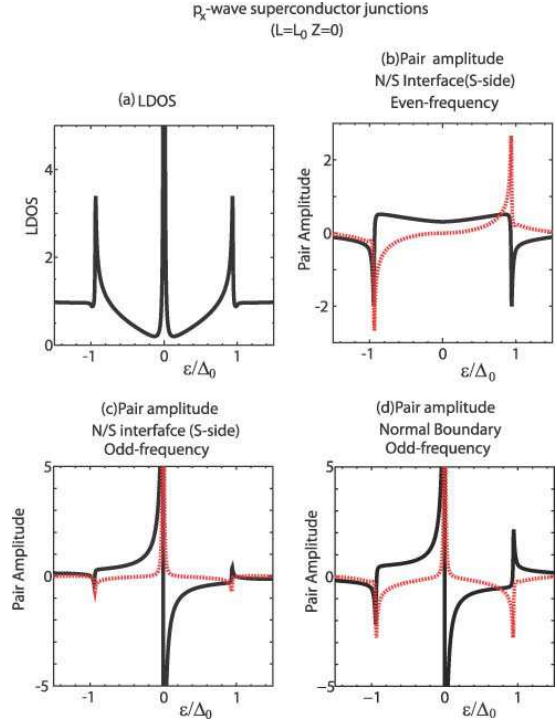}
}
\end{center}
\caption{(Color online) Energy dependence of the LDOS and the pair amplitudes
in $p_{x}$-wave junctions with $L=L_{0}$ and $Z=0$. (a) The LDOS normalized by its value in the normal
state. The solid line: LDOS on the S side of the N/S interface, the dotted
line: LDOS in the N region. Energy dependence of the real (solid line) and
the imaginary (dotted line) part of (b) even-frequency $p_x$-wave pair amplitude on the S
side of the N/S interface, (c) odd-frequency $s$-wave pair amplitude on
the S side of the N/S interface and (d) odd-frequency $s$-wave pair amplitude
at the edge of the N region.}
\label{fig:6}
\end{figure}

In order to get better insight into the spectral property of the
odd-frequency pair amplitude, we calculate the LDOS and the pair amplitudes
as functions of real energy $\varepsilon$. We focus on the even-frequency $%
p_{x}$-wave component of the pair amplitude $C_{1}^{(2)}$ and odd-frequency $%
s$-wave component of the pair amplitude $C_{0}^{(1)}$ at the N/S interface
on the S side and the N boundary. In the N region, the LDOS is independent
of $x$ as shown by Eq. (\ref{LDOS}). For $Z=0$ and $L=L_{0}$ [see Fig.~\ref%
{fig:6}(a)], the LDOS has a zero energy peak (ZEP) due to the formation of
the MARS. There is no jump of the LDOS at the interface since the interface
is fully transparent. The even-frequency pair amplitude at S-side of the N/S
boundary is shown in Fig.~\ref{fig:6}(b). Both the real and imaginary parts
do not vary strongly around $\varepsilon \sim 0$. %

Similar to the case of $s$-wave junctions, the real part of the 
even-frequency component is an even function of $\varepsilon$ 
while the imaginary part is an odd 
function of $\varepsilon$. The corresponding odd-frequency component is
plotted in Fig.~\ref{fig:6}(c). The real (imaginary) part of the pair
amplitude is odd (even) function of $\varepsilon$. The amplitude of the
odd-frequency pair amplitude is enhanced around $\varepsilon = 0$ where the
LDOS has the ZEP. At the edge of the N ($x=-L$), only the odd-frequency
component exists. The line shape of the pair amplitude shown in Fig.~\ref%
{fig:6}(d) is similar to that shown in Fig.~\ref{fig:6}(c).


\begin{figure}[tb]
\begin{center}
\scalebox{0.8}{
\includegraphics[width=9cm,clip]{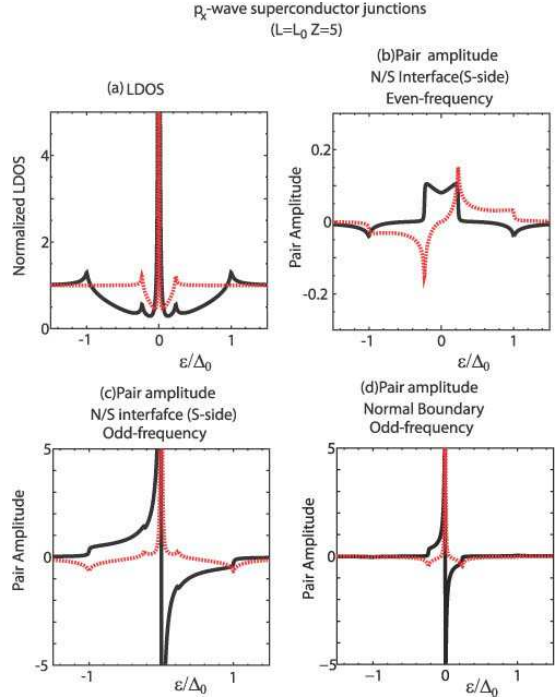}
}
\end{center}
\caption{(Color online) Same as Fig.~\protect\ref{fig:6}, but with $L=L_{0}$ and $Z=5$. }
\label{fig:7}
\end{figure}

In Fig.~\ref{fig:7}, the results of corresponding calculation with $L=L_{0}$ and $Z=5$
are shown. Both the LDOS at the N/S interface and the edge of the N have a
ZEP. In the N region, the LDOS is almost unity due to the absence of the
proximity effect for $| \varepsilon | > 0.24\Delta_{0}$ [see dotted line in
Fig.~\ref{fig:7}(a)].
The LDOS has a ZEP and small peak at $\varepsilon = 0.24\Delta_{0}$.
The corresponding real and imaginary parts of the even-frequency pair amplitude
at the N/S interface also have peaks at this energy (Fig.~\ref{fig:7}(b)]).
The amplitude of the odd-frequency component is enhanced as compared to the
corresponding even-frequency one as shown in Fig.~\ref{fig:7}(c). At the
edge of the N region ($x=-L)$, the amplitude of the odd-frequency component
is almost zero for $\mid \varepsilon \mid > 0.24\Delta_{0}$.
However, around zero
energy, the amplitude of the odd-frequency component is drastically enhanced
as in the case of S-side of the N/S boundary. The penetration of the
odd-frequency component occurs only at low energies.

\begin{figure}[tb]
\begin{center}
\scalebox{0.8}{
\includegraphics[width=9cm,clip]{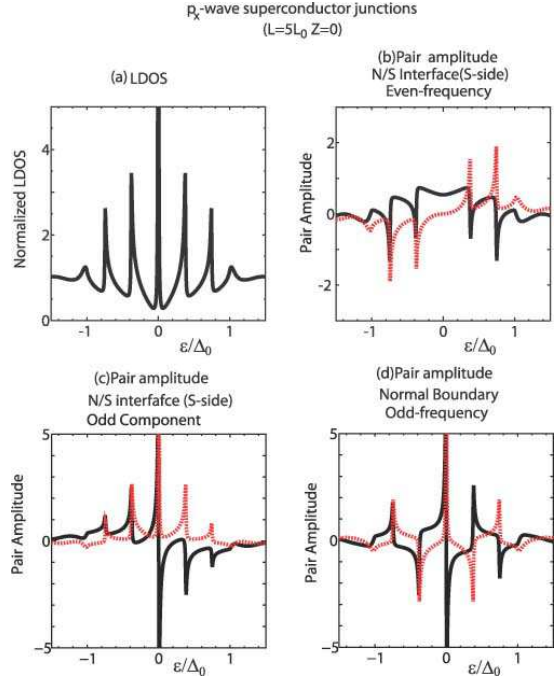}
}
\end{center}
\caption{(Color online) Same as Fig.~\protect\ref{fig:7}, but with $L=5L_{0}$ and $Z=0$. }
\label{fig:8}
\end{figure}

For the longer normal region with $L=5L_{0}$, the resulting LDOS has the ZEP
and a number of peaks at finite energies $\varepsilon$ [see Fig.~\ref{fig:8}(a)]. The even-frequency component of the pair amplitude on the S side of
the N/S boundary also has multiple peaks. The corresponding odd-frequency
component has many peaks with amplitudes strongly enhanced around
$\varepsilon =0$.
Around zero energy, the amplitude of the odd-frequency
component is much larger than that of the even-frequency one
[see Fig.~\ref{fig:8}(c)].
At the edge of the N region, the resulting odd-frequency
component has a significant amplitude as shown in Fig.~\ref{fig:8}(d).
Finally, we focus on the ratio
of the odd- and even-frequency components of the pair amplitude,
$f_{1+}^{(N)}(\varepsilon,\theta)/
f_{2+}^{(N)}(\varepsilon,\theta)$.
\begin{figure}[tb]
\begin{center}
\scalebox{0.8}{
\includegraphics[width=4.5cm,clip]{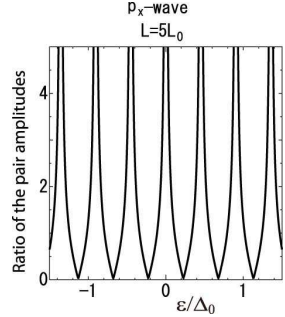}
}
\end{center}
\caption{
Ratio of the pair amplitudes
$f_{1+}^{(N)}(\varepsilon,\theta)/f_{2+}^{(N)}(\varepsilon,\theta)$
as a function of $\varepsilon$ for
$L=5L_{0}$ at the N-side of the N/S interface for $p_{x}$-wave junctions for
$\theta=0$.  }
\label{fig:8-1}
\end{figure}
In Fig. \ref{fig:8-1},
we plot this ratio for $\theta=0$ and $x=0$.
Remarkably, at some energies the odd-frequency pair amplitude exceeds that of the
even-frequency one. In contrast to the $s$-wave case,
there is a huge peak at $\varepsilon=0$ corresponding to the
existence of the MARS. \par

To summarize, we have shown that when the LDOS has a ZEP, the
resulting odd-frequency component is enhanced around $\varepsilon$,
its imaginary part having a ZEP.
It is evident that the odd-frequency pairing state is indispensable
to understand the proximity effect in $p_{x}$-wave superconductor system.

\subsection{$d_{xy}$-wave pair potential}

\begin{figure}[tb]
\begin{center}
\scalebox{0.8}{
\includegraphics[width=20cm,clip]{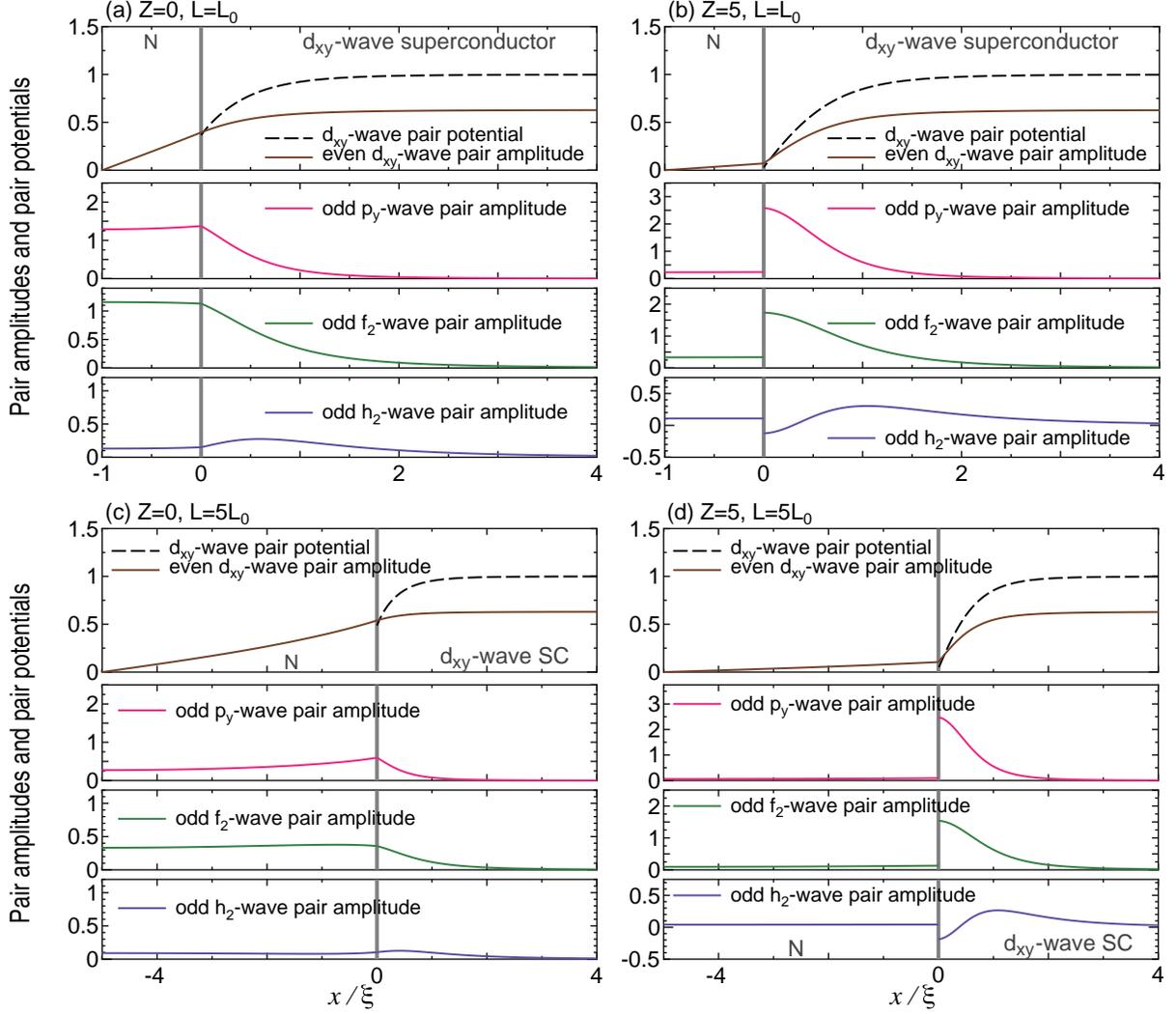}
}
\end{center}
\caption{(Color online) Spatial dependence of the normalized pair potential,
even-frequency and odd-frequency pair amplitudes for $d_{xy}$-wave
superconductor junctions. Here, we choose $\protect\xi=v_{F}/\Delta_{0}$ in
the S region $(x>0)$ and $\protect\xi=L_{0}=v_{F}/2\protect\pi T_{C}$ in the
N region. The pair amplitudes $S_{2}^{(2)}$, $S_{1}^{(1)}$, $%
S_{3}^{(1)}$, and $S_{5}^{(1)}$ are denoted as even $d_{xy}$-wave,
odd $p_{y}$-wave, odd $f_{2}$-wave, and odd $h_{2}$-wave pair amplitudes. 
(a) $Z=0$, $L=L_{0}$, (b) $Z=5$, $L=L_{0}$%
, (c) $Z=0$, $L=5L_{0}$, and (d) $Z=5$, $L=5L_{0}$, respectively. }
\label{fig:9}
\end{figure}

Finally we focus on the $d_{xy}$-wave junctions as shown in
Fig.~\ref{fig:9}. Similar to the above two cases, by changing the length of
the normal region $L$ and the transparency at the interface, we calculate
the spatial dependence of the pair potential and the pair amplitudes in the
Matsubara frequency representation. Here we only concentrate on the lowest
angular momentum of the even-frequency pair amplitude $S_{2}^{(2)}$. The $s$%
-wave component of the pairing amplitude is absent due to the sign change of
the pair potential with respect to the exchange $\theta$ by $-\theta$.
As regards the odd-frequency pair amplitude, we focus on the $S_{1}^{(1)}$, $%
S_{3}^{(1)}$ and $S_{5}^{(1)}$ components where the spatial parity of the
odd-frequency components is odd.
%
These cases correspond to $p_{y}$-wave, $f_{2}$-wave and $h_{2}$-wave
components in Fig.~\ref{fig:9} where the $\theta$ dependence is given by $%
\sin\theta$, $\sin 3\theta$, and $\sin 5\theta$, respectively. In all
cases, the even-frequency component is constant in the S region far away
from the interface and the corresponding odd-frequency components are
absent. The $d_{xy}$-wave pair potential is suppressed at the N/S interface
in all cases. For $Z=5$, the reduction is significant and it is almost zero
at the N/S interface. The even-frequency $d_{xy}$-wave pair amplitude is
reduced towards the N/S interface and monotonically decreases in the N
region similar to the case of $p_{x}$-wave one [see Figs.~\ref{fig:5}(a) and %
\ref{fig:9}(a)]. It does not have a jump at the N/S interface even for
nonzero $Z$ and vanishes at the edge of N region ($x=-L$). On the other hand,
the odd-frequency component is always nonzero at $x=-L$ and has a jump at
the N/S interface. The amplitude of the odd-frequency component is strongly
enhanced near the S-side of the N/S interface. This enhancement is much more
significant for the low transparent interface with large magnitude of $Z$
[see Figs.~\ref{fig:9}(b) and \ref{fig:9}(d)].
On the other hand, for $Z=0$, the odd-frequency components significantly
penetrate into the N region. Note that not only the $p_{y}$%
-wave but also $f_{2}$-wave and $h_{2}$-wave components have the sufficiently large
magnitudes as shown in Figs.~\ref{fig:9}(a) and \ref{fig:9}(c).
The above pair amplitudes are almost constant in the N region.

\begin{figure}[tb]
\begin{center}
\scalebox{0.8}{
\includegraphics[width=9cm,clip]{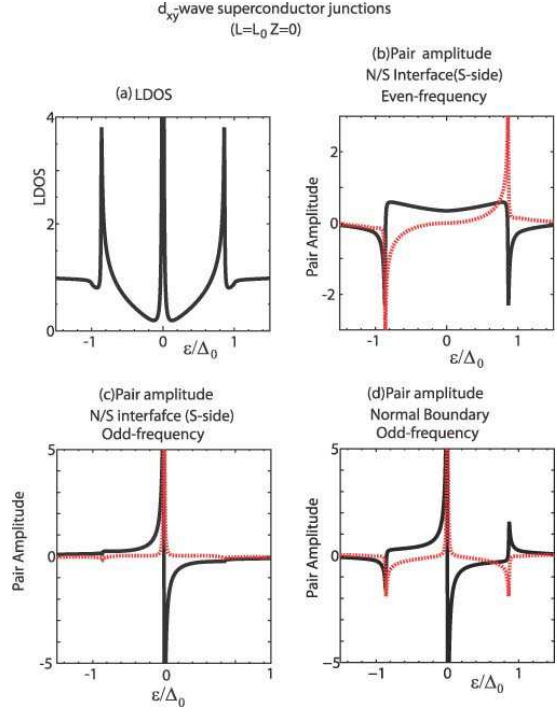}
}
\end{center}
\caption{(Color online) Energy dependence of the LDOS and the pair amplitudes
in $d_{xy}$-wave junctions with $L=L_{0}$ and $Z=0$. (a) The LDOS normalized by its value in the normal
state. The solid line: LDOS on the S side of the N/S interface, the dotted
line: LDOS in the N region. Energy dependence of the real (solid line) and the
imaginary (dotted line) part of (b) even-frequency $d_{xy}$-wave pair amplitude on the S
side of the N/S interface, (c) odd-frequency $p_y$-wave pair amplitude on
the S side of the N/S interface and (d) odd-frequency $p_y$-wave pair amplitude
at the edge of the N region }
\label{fig:10}
\end{figure}

In order to get better insight into the spectral property of the
odd-frequency pair amplitude, we calculate the LDOS and the pair amplitudes
as functions of the real energy $\varepsilon$. We focus on the
even-frequency $d_{xy}$-wave component of the pair amplitude $S_{2}^{(2)}$
and odd-frequency $p_{y}$-wave component of the pair amplitude $S_{1}^{(1)}$ at
the S side of the N/S interface and at the edge of the N region, $x=-L$.
The resulting LDOS has the ZEP due to the formation of the MARS.
Similar to the previously considered cases of $s$-wave and $p_x$-wave junctions,
the LDOS is independent of $x$ as follows from eq. (\ref{LDOS}).
Here, we choose $Z=0$ and $L=L_{0}$ [see Fig.~\ref{fig:10}(a)].
Similar to the $p_{x}$-wave case, the even-frequency pair amplitude
at the N/S boundary on the S-side is shown in Fig.~\ref{fig:10}(b). Both the
real and the imaginary parts do not vary strongly around $\varepsilon \sim 0$.
The real part of the even-frequency component is an even function of $\varepsilon$ while its
imaginary part is an odd function of $\varepsilon$. The corresponding odd-frequency component is plotted in Fig.~\ref{fig:10}(c).
In contrast to the even-frequency component [Fig.~\ref{fig:10}(b)], the real
(imaginary) part of the pair amplitude is an odd (even) function of $%
\varepsilon $. The amplitude of the pair amplitude is enhanced around $%
\varepsilon \sim 0 $ where the LDOS has the ZEP. At the N/S boundary ($x=-L$%
), only the odd-frequency component exists. The line shape of the pair
amplitude as shown in Fig.~\ref{fig:10}(d) is qualitatively
similar to that in Fig.~\ref%
{fig:10}(c). This qualitative behavior of the line shapes is very similar
to that for the corresponding $p_{x}$-wave case.

Similar to the $p_{x}$-wave junction case, when the LDOS has the ZEP, the
resulting odd-frequency component is enhanced around $\varepsilon$. It is
evident that the odd-frequency pairing state is indispensable to understand
the proximity effect in $d_{xy}$-wave superconductor system.

In the following, we comment on the differences between odd-frequency pair
amplitudes in $d_{xy}$-wave and $p_{x}$-wave junctions. In both cases, the
magnitude of the odd-frequency component is enhanced at the interface and in
the normal region. However, the odd-frequency odd-parity state is generated
for $d_{xy}$-wave case, while the odd-frequency even-parity state is
generated for $p_{x}$-wave case. The $s$-wave isotropic component which is
robust against the impurity scattering \cite{Golubov2007}
appears only in the latter case. Then the $d_{xy}$-wave pair amplitude
cannot penetrate into diffusive normal metal while the $p_{x}$-wave one can.
Thus we can naturally understand the presence of proximity effect with the
MARS in $p_{x}$-wave junctions \cite{proximityp,Golubov2007} and its absence
in $d_{xy}$-wave junctions \cite{proximityd,Golubov2007}.

It is instructive to relate the LDOS anomalies in $d_{xy}$-wave and
$p_{x}$-wave junctions to the magnitude of the odd-frequency pairing component.
According to Eq. (\ref{f12}), in $d_{xy}$- and $p_{x}$-wave junctions
the ratio of the magnitude of the odd-frequency
component $f_{1+}^{(N)}(\varepsilon,\theta)$
to the even-frequency one $f_{2+}^{(N)}(\varepsilon,\theta)$  is

\begin{equation}
\frac{\left\vert f_{1+}^{(N)}(\varepsilon, \theta)\right\vert }{\left\vert
f_{2+}^{(N)}(\varepsilon, \theta)\right\vert }=\frac{\left\vert 1/t+t\right\vert }{\left\vert
1/t-t\right\vert }=\left\vert {\rm cotan} \left( \frac{2\varepsilon }{v_{Fx}}%
(L+x)\right) \right\vert.  \label{f12_E}
\end{equation}

It follows from the above expression that at the edge of the N region, $x=-L$,
the odd-frequency component dominates at all energies. On the other hand,
at the N/S interface, $x=0$, the odd-frequency component dominates at energies
$\varepsilon =\varepsilon _{n}$ corresponding to the LDOS has peak positions
\begin{equation}
\varepsilon _{n}=\frac{\pi v_{Fx}n}{2L}, \text{ \ \ }n=0,1,2,...
\label{f12_pd}
\end{equation}
For $n=0$ Eq. (\ref{f12_pd}) describes the mid-gap Andreev bound state
and higher $n$ correspond to the subgap resonances for large N region thickness.
Therefore, we can conclude that in $d_{xy}$-wave and $p_{x}$-wave junctions the odd-frequency
component dominates over the even-frequency one at the N/S interface at the energies
corresponding to the LDOS peak positions, including the prominent zero-energy peak (MARS).
Moreover, the odd-frequency component always dominates at the edge of the N region, $x=-L$,
where the breaking of translational invariance is the strongest because of sign change of the
pair amplitude at that point.

\section{Conclusions}

In summary, using the quasiclassical Green's function formalism, we have
shown that the odd-frequency pairing state is ubiquitously generated in the
normal metal/superconductor (N/S) ballistic junction system, where the 
length of the normal region is finite. It is shown that the even-parity 
(odd-parity) pair potential in the superconductor induces the odd-frequency 
pairing component with spin-singlet odd-parity 
(spin-triplet even-parity). As regards the 
symmetry of the superconductor, we have chosen typical three cases, 
spin-singlet $s$-wave, spin-triplet $p_{x}$-wave and spin-singlet $d_{xy}$%
-wave. In the latter two cases, mid gap Andreev resonant state (MARS)
appears at the N/S interface. Even for conventional $s$-wave junctions, the
amplitude of the odd-frequency pairing state is enhanced at the N/S
interface with fully transparent barrier. By analyzing the spectral
properties of the pair amplitudes, we found that the magnitude of the
resulting odd-frequency component at the interface can exceed that of the
even-frequency one. For the case of $p_{x}$-wave and $d_{xy}$-wave
junctions, the magnitude of the odd-frequency component at the S side of the
N/S interface is significantly enhanced. The magnitude of the induced
odd-frequency component is enhanced in the presence of the midgap Andreev
resonant state due to the sign change of the anisotropic pair potential at
the interface. The LDOS has a zero energy peak (ZEP) both at the interface
and in the N region. At the edge of the N region, only the odd-frequency
component is non-zero.

The underlying physics behind these phenomena is related to the breakdown of
translational invariance near the N/S interface where the pair potential
$\bar{\Delta}(x)$ acquires a spatial dependence. As a result, an
odd-frequency component is quite generally induced near the interface. The
breakdown of translational invariance is the strongest when the pair potential
%
changes sign upon reflection like in the case of $p_{x}$-wave and $d_{xy}$-wave
junctions, then the magnitude of odd-frequency component is the largest.
Moreover, the phase of the interface-induced odd-frequency component has a
$\pi /2$ shift from that in the bulk of S. Therefore, as shown above, the
odd-frequency component $f_{1\pm }^{(i)}(\omega _{n},\theta )$ becomes
purely imaginary quantity and the peak structure in the LDOS naturally follows
from the normalization condition.

We have also shown that in the N/S junctions with $s$-wave superconductors the
classical McMillan-Rowell oscillations \cite{Rowell} can also be reinterpreted
in terms of odd-frequency pairing. As follows from Eq.(\ref{f12_sgp}), at
the energies corresponding to the subgap peaks in the N/S junction, the
odd-frequency component dominates over the even-frequency one. This is
remarkable application of the odd-frequency pairing concept when one can
re-interpret the well-known resonance structure.

In the present study, we clarified the details of proximity effect of the
odd-frequency pairing state induced at the N/S boundary. We reinterpreted
the appearance of the MARS in terms of the enhanced odd-frequency pair
amplitude. Though we explicitly studied the N/S junctions only, the
odd-frequency pairing state is also expected near impurities and within
Abrikosov vortex cores, where the amplitude of the pair potential is
reduced. The present result indicates \textit{the ubiquitous presence of
odd-frequency pairing states} because most of real superconductors are not
uniform. That means that the odd-frequency pairing is not at all a rare
situation as was previously assumed. Thus we believe that the odd-frequency
pairing may become an important concept in understanding the physics of
non-uniform superconducting systems.

In the present paper, the proximity effect is studied in the ballistic
limit. In the present case, the enhanced odd-frequency pair amplitude appears
in the N region both for $p_{x}$-wave and $d_{xy}$-wave junctions. It is
very interesting to study in the intermediate regime \cite{Lofwander} since
the parity of these states are different. In the diffusive limit, the
proximity effect survives only for the former case. There are several
interesting phase coherent effects relevant to MARS \cite{Josephson,Meissner}%
. These preexisting phenomena can be reinterpreted in terms of the
odd-frequency paring state.

One of the authors Y.T. expresses his sincerest gratitude to discussions
with M. Ueda and S. Kashiwaya and K. Nagai.
This work is supported by Grant-in-Aid for Scientific Research (Grant No.
17071007 and 17340106) 
from the Ministry of Education, Culture, Sports, Science and Technology of
Japan, by Japan Society for the Promotion of Science (JSPS) and by NanoNed project TCS.7029.
The computation in this work was done using the facilities of the Supercomputer Center,
Institute for Solid State Physics, University of Tokyo. This work is supported by NTT
basic research laboratory.


\end{document}